\documentclass[preprint,preprintnumbers,amsmath,amssymb]{revtex4}

\usepackage{graphicx}
\usepackage{dcolumn}
\usepackage{bm}
\usepackage{subfigure}

\begin{document}

\author{Alexandre E. Hartl}
 \altaffiliation[ ]{Department of Mechanical and Aerospace Engineering, North Carolina State University, Raleigh, NC 27695-7910.}
 
\author{Bruce N. Miller}
 \altaffiliation[ ]{Department of Physics and Astronomy, Texas Christian University, Forth Worth, TX 76129.} 

\author{Andre P. Mazzoleni}
 \altaffiliation[ ]{Department of Mechanical and Aerospace Engineering, North Carolina State University, Raleigh, NC 27695-7910.}

\date{\today}

\title{Dynamic Modeling and Simulation of a Real World Billiard}

\begin{abstract}

Gravitational billiards provide an experimentally accessible arena for testing formulations of nonlinear dynamics. We present a mathematical model that captures the essential dynamics required for describing the motion of a realistic billiard for arbitrary boundaries. Simulations of the model are applied to parabolic, wedge and hyperbolic billiards that are driven sinusoidally.  Direct comparisons are made between the model's predictions and previously published experimental data. It is shown that the data can be successfully modeled with a simple set of parameters without an assumption of exotic energy dependence. 

\end{abstract}

\maketitle
Gravitational billiards provide an experimentally accessible arena for testing formulations of nonlinear dynamics. One and two-dimensional Hamiltonian versions have long provided easily visualized systems that exhibit a wide range of stable and chaotic behavior \cite{Barroso:1,Miller:1,Korsch:1,Miller:2}. The system consists of a particle undergoing elastic collisions within a rigid boundary, where the particle follows a ballistic trajectory between collisions. When the boundary is periodically driven, Fermi acceleration may result \cite{Lenz:1}, establishing a connection with cosmic ray physics and cosmology. Recent interest in dynamics has been focused on dissipative systems such as granular media. While inelasticity in these systems is usually represented by a collisional restitution coefficient, it has been observed that the inclusion of rotational friction induces qualitative changes in behavior \cite{Ben-naim:1,Orlando:1} that cannot be explained by other means. Similarly, in billiard experiments \cite{Olafsen:1}, when friction is left out of the formulation, it appears that one is forced to make unphysical assumptions about the source of energy loss to approximately replicate the experimental data \cite{Gorski:1}. This paper considers the more realistic situation of an inelastic, rotating, gravitational billiard in which there are retarding forces due to air resistance and friction.  In this case the motion is not conservative, and the billiard is no longer a particle, but a sphere of finite size. We present a mathematical model that captures the relevant dynamics required for describing the motion of this "real world" billiard for arbitrary boundaries.  The model is applied to parabolic, wedge and hyperbolic billiards that are driven sinusoidally.  Direct comparisons are made between the model's predictions and experimental data previously collected \cite{Olafsen:1}.  Although several studies have investigated the effect of variable elasticity in relation to the gravitational billiard, this study is the first to incorporate rotational effects and additional forms of energy dissipation.  

\begin{figure}[htb]
  \begin{center}
    \leavevmode
      \includegraphics[width=3.0in]{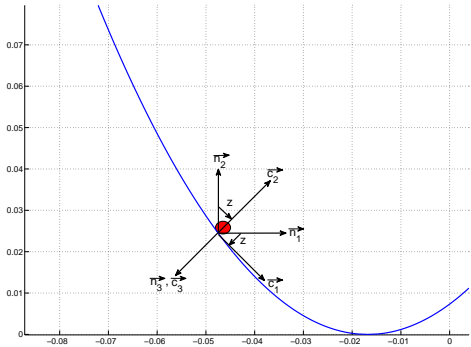}
    \caption{The coordinate frame at the collision point is denoted by the $\mathbf{c}$$-$frame.  Reference frames $\mathbf{c}$ and $\mathbf{n}$ are related to each other by angle $z$.} 
    \label{fig:n_frame2}
  \end{center}
\end{figure}

\begin{figure}[htb]
\begin{center}
\subfigure{\includegraphics[scale=0.37]{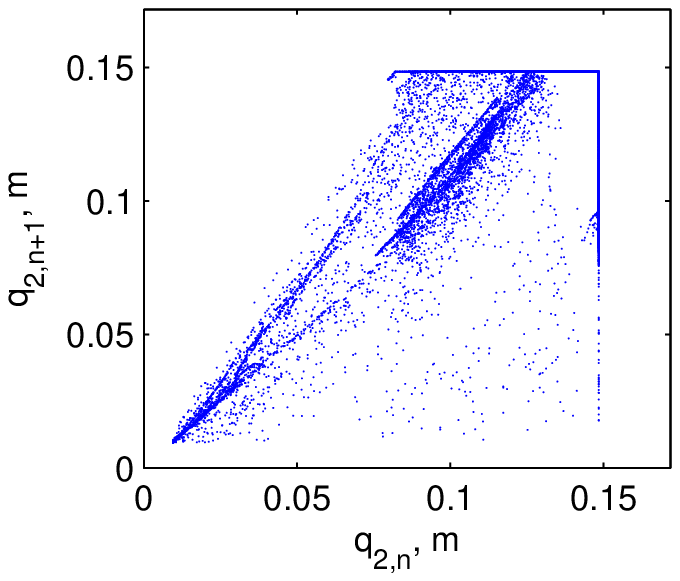}}
\subfigure{\includegraphics[scale=0.37]{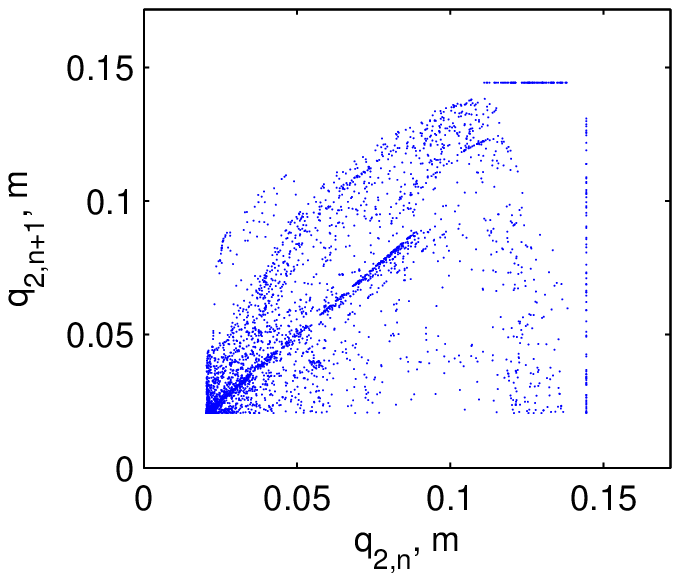}}
\subfigure{\includegraphics[scale=0.37]{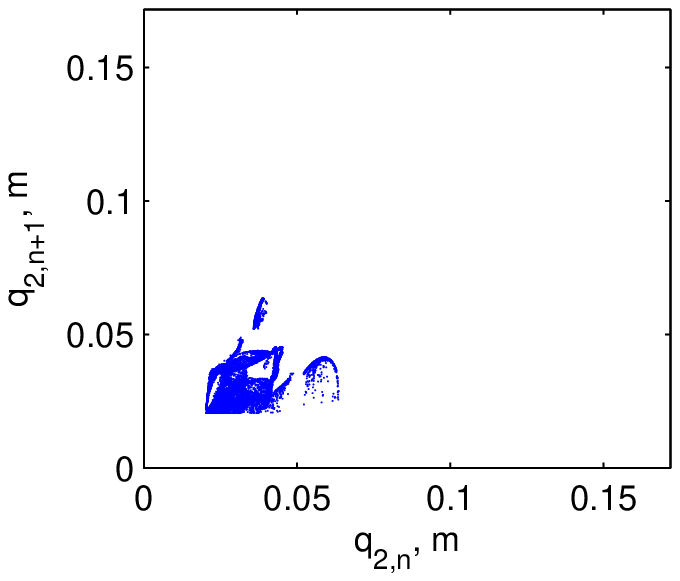}}\\ 
\subfigure{\includegraphics[scale=0.37]{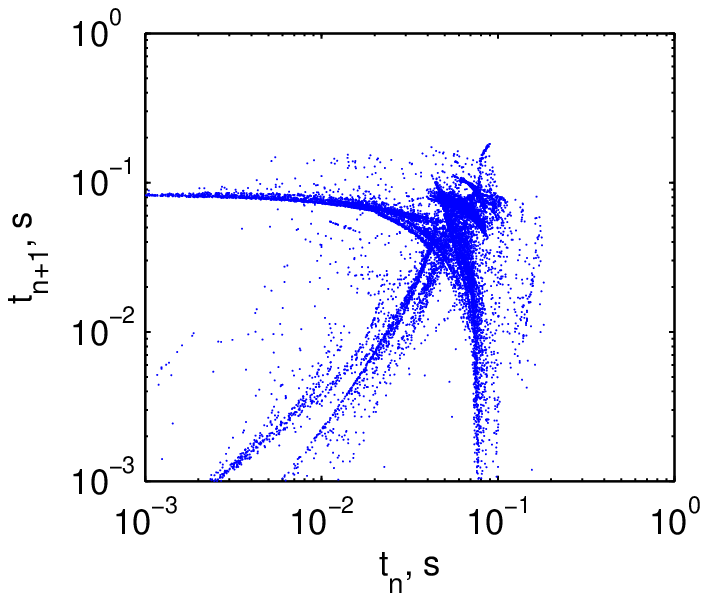}} 
 \subfigure{\includegraphics[scale=0.37]{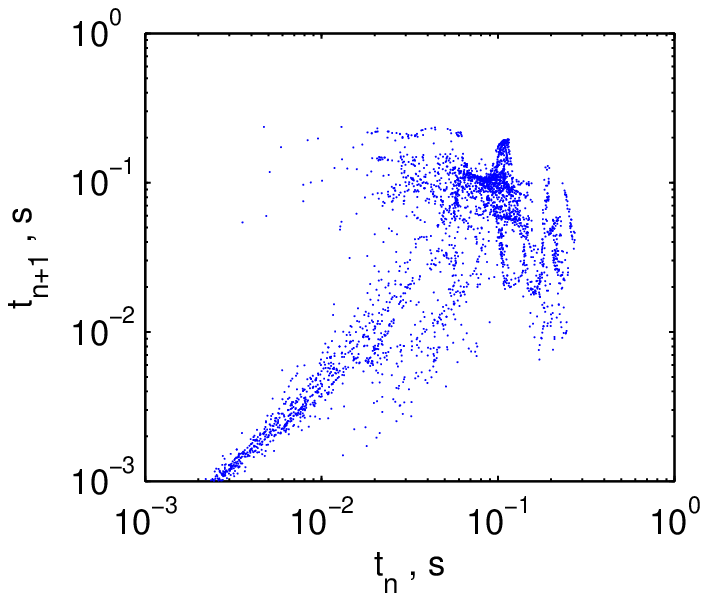}} 
\subfigure{\includegraphics[scale=0.37]{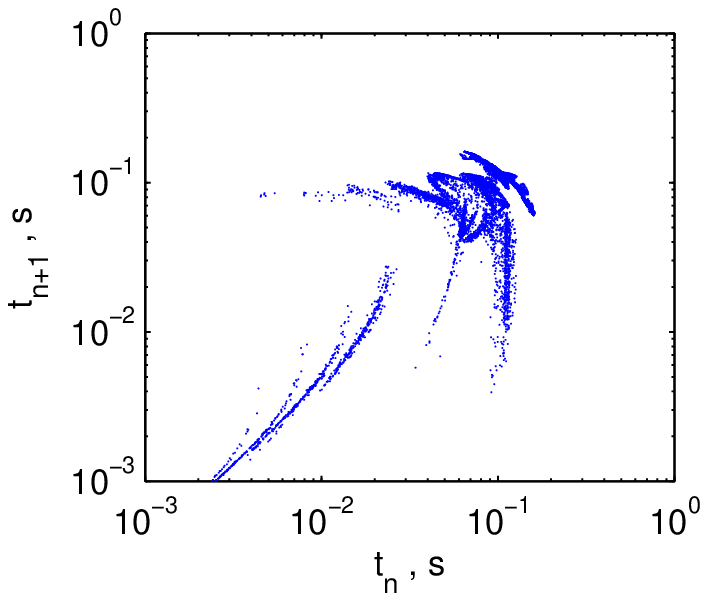}}
\end{center}
\caption{The spatial and temporal mappings of the collision heights (top row) and times of flight (bottom row).  From left to right: the wedge at 6.6 Hz, the hyperbola at 5.8 Hz and the hyperbola at 4.5 Hz.}
\label{fig:2} 
\end{figure}

\begin{figure}[htb]
\begin{center}
\subfigure{\includegraphics[scale=0.37]{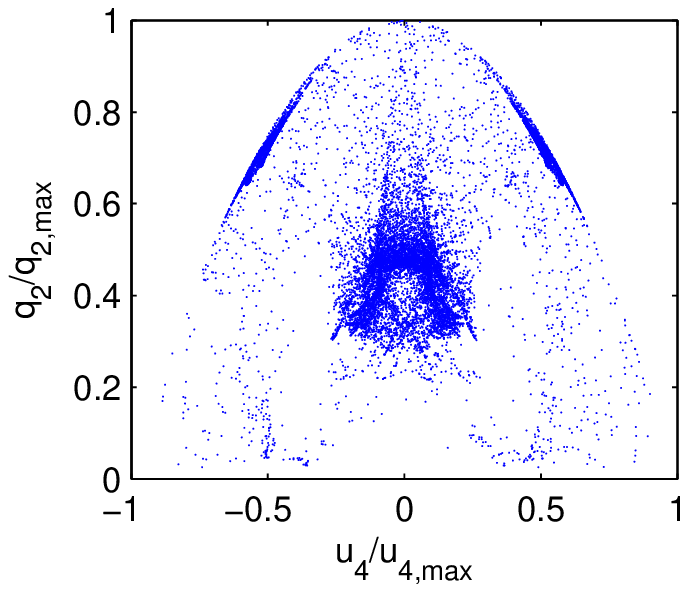}}
\subfigure{\includegraphics[scale=0.37]{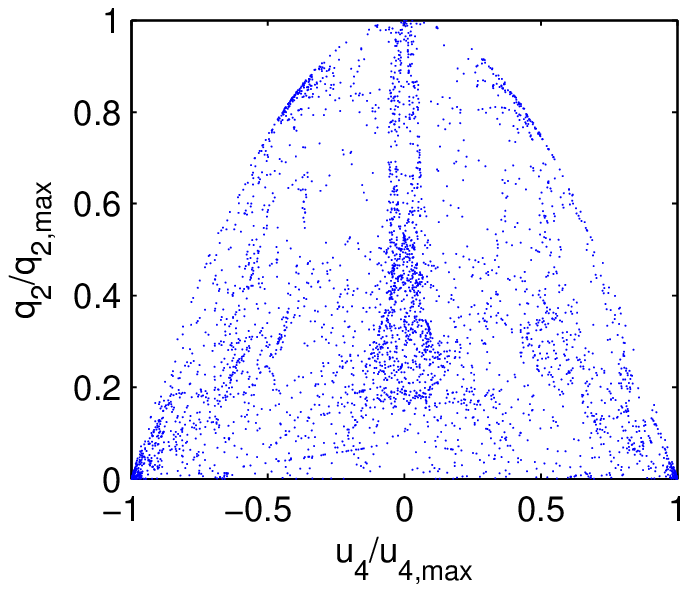}}
\subfigure{\includegraphics[scale=0.37]{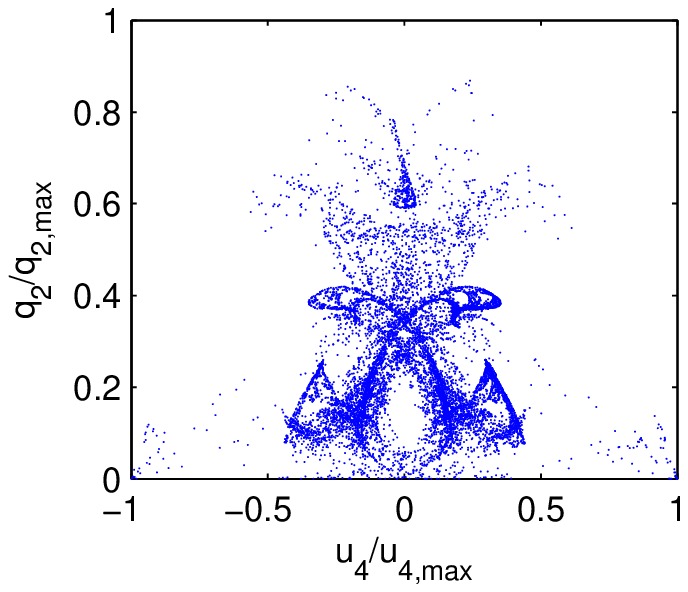}} 
\end{center}
\caption{The normalized collision heights $q_{2}/q_{2,max}$ versus the normalized tangential velocity $u_{4}/u_{4,max}$.  Starting from left to right: the wedge at 6.6 Hz, the hyperbola at 5.8 Hz and the hyperbola at 4.5 Hz. }
\label{fig:3} 
\end{figure}

The ergodic properties of Hamiltonian gravitational billiards are well studied \cite{Miller:1,Korsch:1,Miller:2,Wojt:1}. It has been shown that the parabolic billiard is completely integrable having stable, periodic orbits \cite{Korsch:1}. Studies of the wedge billiard demonstrated that the billiard's behavior depends on the vertex angle, defined as $2\theta$ \cite{Miller:1}.  For 0 $<\theta<45^{o}$, the phase space contains coexisting stable and chaotic behavior. For $\theta = 45^{o}$ the motion is completely integrable, while for $\theta>45^{o}$, the motion is chaotic.  These results have been confirmed through experiments for an optical billiard with ultra cold atoms \cite{Milner:1}.  The motion of a hyperbolic billiard exhibits characteristics of the parabolic billiard for low energy, where the motion is near the origin, and wedge billiard at high energy, where the motion is mostly concentrated at its asymptotic limits \cite{Miller:2}. 
\\
\indent
Feldt and Olafsen\cite{Olafsen:1} have experimentally studied the real world billiard through several rounds of testing.  One experiment consisted of a steel ball moving within a closed reflective aluminum boundary shaped either as a parabola, wedge or hyperbola.  The container was driven in the horizontal direction to compensate for energy losses resulting from collisions.  Imaging software determined the ball's position and velocity at the collision points.  The study results indicated regular motion for the parabola and chaotic motion for the wedge; the motion for the hyperbola was found to be frequency dependent, sharing characteristics of the parabola at low-driving frequencies and the wedge for higher-driving frequencies.  
\\
\indent
In this work direct comparisons are made between simulations of the model system and the experimental data of Feldt and Olafsen. To date, G$\acute{o}$rski and Srokowski\cite{Gorski:1} are the only investigators known to have theoretically studied the experiments conducted by Feldt and Olafsen.  In their model they consider an inelastic, gravitational billiard for the case of no friction or rotation, and no drag. In order to replicate the main features of the experiments, it was necessary to resort to a surprising, unconventional representation of the restitution coefficient energy dependence.
\\
\indent
For the general billiard-boundary system, the laboratory frame is defined by three mutually perpendicular unit vectors ($\mathbf{n_{1}},\mathbf{n_{2}},\mathbf{n_{3}}$), where $\mathbf{n_{2}}$ is perpendicular to the plane formed by vectors $\mathbf{n_{1}}$ and $\mathbf{n_{3}}$.  The reference frame at the collision point between the billiard and boundary is defined by the $\mathbf{c}$$-$frame, and is related to the laboratory frame by a translation and rotation of coordinates.  Figure~\ref{fig:n_frame2} shows the $\mathbf{c}$ and $\mathbf{n}$$-$frames for the two-dimensional case, where the frames are related by angle $z$.  For curved boundaries angle $z$ varies along the curve, and is uniquely determined for each collision.  
\\
\indent
In the following we consider driven parabolic, wedge and hyperbolic boundaries defined mathematically (in the laboratory frame), respectively, as: 
\begin{eqnarray}
\label{1033:u}
&& q_{2} = f(q_{1}) = a\left(q_{1} - \Delta q_{1}\right)^{2} + c   \\
&& q_{2} = f(q_{1}) = b\left|q_{1} - \Delta q_{1}\right| + c  \\
&& q_{2} = f(q_{1}) = \sqrt{a\left(1 + \beta \left(q_{1} - \Delta q_{1}\right)^{2}\right)} - \delta \ 
\end{eqnarray}
where $a = 0.26 cm^{-1}$, $b = 1.85$, $c = 0.63 cm$, $\alpha = 40.3 cm^{2}$, $\beta = 0.08 cm^{-2}$ and $\delta = 4.45 cm$. These are the values used in the experiments of Feldt and Olafsen. The boundaries oscillate horizontally and their position at time $t$ is defined by 
\begin{equation}
\label{1034:functionc}
\Delta q_{1}\left(t\right) = A sin \omega t
\end {equation}
where $A$ is the amplitude and $\omega = 2\pi f$ is the oscillation angular frequency in rad/sec, and $f$ is the oscillation frequency in hertz.  Experimentally, note that the boundaries are sealed off by a top and thin pieces of Plexiglas on the sides, rendering the system to two-dimensions.  
The billiard's position is tracked in time by following its geometric center, where its position is defined by 
\begin{equation}
\label{1011:eq}
\mathbf{q}=q_{1}\mathbf{n_{1}}+q_{2}\mathbf{n_{2}}+q_{3}\mathbf{n_{3}}
\end {equation}
For the numerical simulations presented in this study, $q_{3} = 0$.  In reality, however, $q_{3} \neq 0$ because of slight chattering in the $\mathbf{n_{3}}$ direction.  Private communications with an author of \cite{Olafsen:1} reveal that the system's noise induced by this effect is small.  The actual billiard considered in the simulations and used in the experiments is a $3.1$ mm diameter steel chrome ball weighing approximately $0.13028$ g $+/-$ $3\times10^{-5}$ g.  
\\
\indent
In this study we employ a modified version of the impact theory set forth by Kane and Levinson for collisions between the billiard and boundary\cite{Kane:1}, where we account for the effects of a moving boundary.  The original theory provides a direct method for computing the billiard's generalized speeds post collision considering fixed boundaries.  Kane and Levinson make the following assumptions in their model: first, the contact area between the objects is a single point through which all forces are exerted.  Second, the total collision impulse is represented by the integral of the forces over the entire collision time.  Third, the coefficients of restitution, static friction and kinetic friction are constants to be determined experimentally.  The theory initially assumes no slipping at the contact point between the billiard and boundary.  A set of values for the generalized speeds are generated, and are valid if and only if the no-slip condition is satisfied.  If the no-slip condition is violated, then a new set of values for the generalized speeds are developed under the assumption of slipping.  Refer to reference \cite{Kane:1} for a detailed derivation of the theory. 
\\
\indent
Between collisions we make use of a trajectory model that numerically tracks the billiard's motion after each "bounce".  The model is used to reinitialize the generalized speeds at the point of initial contact with the boundary.  For a billiard moving through air, its motion is affected by gravity, air resistance (drag) and additional aerodynamic forces due to its spinning motion (the Magnus effect).  Here we use the drag force specified by Fowles\cite{Fowles:1}, which considers both the linear and quadratic components of the drag force. The Magnus Force is neglected since its overall effect is small.  Between boundary encounters the trajectory equations consist of second-order, nonlinear, coupled differential equations which are solved numerically by using a fourth-order Runge-Kutta method.  The procedure for detecting collisions between the billiard and boundary works by locating the minimum distance between the objects at each time step, and comparing that distance to a specified tolerance, which for our case is the billiard's radius $b$.  If the distance is less than or equal to the tolerance, then a collision is reported.  Otherwise, it is concluded that no collision has occurred.  Application of the above procedure results in solving a second-order algebraic equation for the wedge, a third-order algebraic equation for the parabola and a fourth-order algebraic equation for the hyperbola.  We determine the roots of the equations by means of the Newton-Raphson method.  If a collision is detected, the collision time and collision location are approximated by interpolation methods. The procedure for finding the collision time is based on a method due to Baraff \cite{Baraff:1}. 
\\
\indent
Collisions with the boundary result in energy losses stemming from the restitution in the normal direction and friction in the transverse direction.  Feldt and Olafsen \cite{Olafsen:1} suggest a coefficient of restitution of 0.9 between the steel billiard and the aluminum boundary, noting that its value is velocity dependent.  However, as we see see in the following, the coupling between the normal and tangential contact forces during the impact reduces this coefficient considerably. Further, it has been observed that the coefficient of restitution depends on the incident angle at impact \cite{Louge:1}.  Kane and Levinson's impact theory remains practical when it is supplemented by experimental measurements capturing the coefficients of restitution and friction.  Toward this end the parabolic billiard is used as a standard test case for establishing the coefficient of restitution since experiments have shown that it exhibits stable, period-one orbits.  The experiment described in \cite{Olafsen:1} resulted in an orbit height of approximately $0.075$ m; an apparent value for the coefficient of restitution is estimated by matching the orbit height of the simulation to the experiment.  First, note that for steel on aluminum, experiments reveal that the coefficient of static and kinetic friction is approximately $0.61$ and $0.47$ \cite{eng:1}, respectively.  If the numerical model applies a coefficient of restitution of $e = 0.393$ along with the friction coefficients specified above, then the simulation replicates the experiment; as a result we apply this $e$ value for all boundary shapes considered in this paper. If sufficient energy is supplied to the system, the billiard's trajectory eventually mode-locks into a stable period-one orbit. The orbit moves up or down the parabola as the driving frequency is increased or decreased, respectively.  If insufficient energy is given to the system, then the parabola will explore multiple trajectories.
\\
\indent
In granular media simulations, a method of preventing inelastic collapse of particles is to set the coefficient of restitution to its elastic limit of $1$ if collisions occur too frequently \cite{Luding:1}.  Additionally, studies have demonstrated that the coefficient of restitution approaches a value of $1$ as the normal component of the impact velocity approaches $0$.  As a consequence we apply a coefficient of restitution of $1$ in our simulations if the relative velocity (between the billiard and boundary) in the normal direction is sufficiently small that it results in inelastic collapse, where a ``nearly infinite'' number of collisions occur in a finite time \cite{Luding:1}.  In practice, this assumption is only applied at the start of simulations (to initiate motion) and for brief instances of time during the simulation (as explained above).  
\\
\indent
Each simulation tracks a single trajectory consisting of 25,000 billiard-boundary collisions.  The billiard is initially at rest, but is quickly propelled into the air by the energy transmitted from the boundary to the billiard.  The collision height $q_{2}$ and time $t$ between consecutive bounces are extracted from the simulations and are shown in Figure~\ref{fig:2} for multiple boundaries driven at varying frequencies.  The successive mappings of the collision heights and times of flight show exceptional resemblance to the experimental data even though a constant coefficient of restitution is used.  In reality, the coefficient of restitution is velocity dependent.  The numerical model, however, detects additional collisions (i.e., collisions that occur in rapid succession) not reported by the experiments.  This is due to the lower resolution of the experimental data.  As a result the numerical model observes more collisions at both the small and large values of the height $q_{2}$; thus the time mappings have longer time tails associated with shorter times of flight for these collisions.
\\
\indent
The plot of the wedge driven at 6.6 Hz reveals that the motion appears chaotic as suggested by the experiments. The billiard is continuously driven to the top of the wedge, with most of the collision points laying above the $q_{2,n} = q_{2,n+1}$ line in the return map.  The time map also shows indications of chaotic behavior and a similar concentration of points to the experimental data.  The hyperbola driven at 5.8 Hz resembles the unstable behavior of the wedge in both position and time, and shows a likeness to the experiment.  At 4.5 Hz, the billiard's motion is confined to the regions near the hyperbola's vertex, and shows the semblance of a regular pattern not noted in the experimental data due to possible smearing of the data as seen in Figure~\ref{fig:2}.  Patterns are also detected in the temporal mapping of the hyperbola at this driving frequency.  
\\
\indent
Following the experiments the phase space is further investigated by plotting the normalized collision height $q_{2}/q_{2,max}$ versus the normalized tangential velocity $u_{4}/u_{4,max}$ post collision.  The quantities $q_{2,max}$ and $u_{4,max}$ are defined as the maximum energy values that the billiard can possess at the collision height $q_{2}$ if all the energy were potential or kinetic, respectively.  A completely stable period-one orbit for a perfectly elastic billiard is characterized by having zero tangential velocity assuming collisions with a symmetric boundary.  For the parabolic billiard driven at 5.4 Hz, the numerical model predicts a small value for the normalized tangential velocity when the billiard achieves a stable period-one orbit; the mapping is a single point that has the value $\left(q_{2}/q_{2,max},u_{4}/u_{4,max}\right)$ =  $\left(0.376, \pm 0.0372\right)$.  The experimental data, however, shows that the normalized tangential velocity is a thin band about zero, where the range in velocity and height is caused by the noise in the system and small variations in the coefficient of restitution.  Figure~\ref{fig:3} shows the results for the remaining boundaries. For the case of the wedge, the billiard explores much of the phase space; the hyperbolic billiard driven at the higher frequency exhibits similar behavior to the wedge, but examines more of the phase space.  For both shapes there are regions that have concentrations of points not indicated by the experiments.  For the hyperbolic billiard driven at the lower frequency, regular patterns develop. 
\\
\indent
Note that the plots in Figure~\ref{fig:3} represent a significant deviation from the results indicated by the experiments.  The difference is qualitative and is not explained by the extra collisions reported by the numerical simulations.  Potentially, the source of the difference may lay in the finite resolution of the imaging software, resulting in measurement uncertainty, where the normalized collision height and normalized tangential velocity are calculated quantities that depend on the accurate resolution of the billiard's position, velocity and velocity components at the collision points.  
\\
\indent
\\
\indent
\\
\indent
There are open questions concerning how best to model impacts between systems of solid objects, such as granular media \cite{Orlando:1}. Examining the ergodic properties of a gravitational billiard provides an experimentally accessible arena for testing and comparing a variety of impact models. Here we have presented one model that captures the relevant dynamics required for describing the motion of a real world billiard for arbitrary boundaries.  The model considers the  realistic situation of an inelastic, rotating, gravitational billiard in which there are retarding forces due to air resistance and friction.  We have used the model to investigate driven parabolic, wedge and hyperbolic billiards, and demonstrated that the parabola has regular motion, while the wedge and hyperbola (at high driving frequencies) appeared chaotic.  The hyperbola, at low driving frequencies, also showed regular motion.  The simple representation of the coefficient of restitution employed in the model resulted in good agreement with the recent experimental data of Feldt and Olafsen for all boundary shapes \cite{Olafsen:1}, but not for secondary quantities derived from the data. The model also predicted additional collisions not detected by the data. The coefficient of restitution introduces the most uncertainty in modeling the billiard-boundary system, and resolution of this problem will require additional experiments. It is interesting that the optimum numerical value is very different if rotation induced friction is not included. We will pursue this surprising effect in a longer work.
\\
\indent
This research was supported by a grant from Frank Ziglar, Jr. of North Carolina State University.  The authors would like to thank Dr. Jeff Olafsen of Baylor University for his helpful interactions.
\\

\end{document}